\documentclass[pra,twocolumn,showpacs,preprintnumbers,amsmath,amssymb]{revtex4-1}

\usepackage{graphicx}
\usepackage{dcolumn}
\usepackage{bm}
\usepackage{color}

\usepackage{enumerate}

\def\rr{{\bm r}}
\def\nn{{\bm n}}
\def\kk{{\bm k}}
\def\non{\nonumber}
\def\xisk{{\xi_{\rm SkX}}}
\def\br{\rho}
\def\tk{\tilde{k}}
\def\ta{\tilde{a}}
\def\td{\tilde{d}}
\def\txi{\tilde{\xi}}
\def\tl{\tilde{l}}

\def\tE{\tilde{E}}
\def\AJ{\mathcal{A}_J}
\def\AD{\mathcal{A}_D}
\def\AB{\mathcal{A}_B}
\def\BJ{\mathcal{B}_J}
\def\BB{\mathcal{B}_B}

\usepackage{}	
\begin{document}

\preprint{APS/123-QED}


\title{Skyrmionic magnetization configurations at chiral magnet/ferromagnet heterostructures}
\author{Yuki Kawaguchi$^1$, Yukio Tanaka$^2$, Naoto Nagaosa$^{1,3}$}
\affiliation{
$^1$Department of Applied Physics, University of Tokyo, Tokyo 113-0033, Japan\\
$^2$Department of Applied Physics, Nagoya University, Nagoya 464-8603, Japan\\
$^3$RIKEN Center for Emergent Matter Science (CEMS), Wako, Saitama 351-0198, Japan
}
\date{\today}

\begin{abstract}
We consider magnetization configurations at chiral magnet (CM)/ferromagnet (FM) heterostructures.
In the CM, magnetic skyrmions and spin helices emerge due to the Dzyaloshinskii-Moriya interaction, which then penetrate into the adjacent FM. 
However, because the non-uniform magnetization structures are energetically unfavorable in the FM, the penetrated magnetization structures are deformed,
resulting in exotic three-dimensional configurations, such as skyrmion cones, sideways skyrmions, and twisted helices and skyrmions.
We discuss the stability of possible magnetization configurations at the CM/FM and CM/FM/CM hybrid structures within the framework of the variational method, 
and find that various magnetization configurations appear in the ground state, some of which cause nontrivial emergent magnetic field.
\end{abstract}

\pacs{75.70.Cn, 75.30.-m, 75.30.Kz, 72.25.Ba}



\maketitle

\section{Introduction}
Magnetic skyrmions are topologically protected non-coplanar configurations of magnetization.
In contrast to vortices and monopoles, 
they can be embedded in a uniform magnetization configuration and behave as particle-like objects~\cite{Skyrme}.
Since their first observation in the metallic chiral magnet (CM) MnSi
by neutron scattering~\cite{Muhlbauer2009} and in (Fe, Co)Si by Lorentz transmission electron microscopy~\cite{Yu2010},
properties of magnetic skyrmions have been extensively investigated~\cite{NagaosaTokura2013}.
The emergent electromagnetic field generated by skyrmionic configurations changes the transport property of the conduction electrons,
resulting in the topological Hall effect~\cite{Binz2008,Lee2009,Neubauer2009,Kanazawa2011,Huang2012,Li2013} 
and the electromagnetic induction~\cite{Schulz2012}.
The coupling between the magnetization and conduction electrons also enables us
to control the motion of skyrmions by an electric current~\cite{Jonietz2010,Yu2012, Iwasaki2013a, Iwasaki2013b}:
Due to the topological nature of skyrmions, they robustly survive in dynamics,
and the mobility is much higher than that of magnetic domains and helical configurations.

Magnetic skyrmions are observed in CMs, such as
metallic MnSi~\cite{Muhlbauer2009,Pappas2009,Pfleiderer2010J}, Fe$_{1-x}$Co$_x$Si~\cite{Munzer2010,Yu2010}, MnGe~\cite{Yu2011}, Fe$_{1-x}$Mn$_x$Ge~\cite{Shibata2013}, and insulating Cu$_2$OSeO$_3$~\cite{Adams2012,Seki2012S}.
These materials have non-centrosymmetric B20-type crystal structures,
where the relativistic Dzyaloshinskii-Moriya (DM) interaction~\cite{Dzyaloshinskii1958, Moriya1960} 
stabilizes a crystalline structure of skyrmions, called a skyrmion crystal (SkX), under an external magnetic field.
Besides the CMs, a lattice of atomic-scale skyrmions is observed in an Fe monolayer on Ir(111) using the spin-polarized scanning tunneling microscopy~\cite{Hinze2011},
where the four-spin interaction, as well as the DM interaction coming from the strong spin-orbit coupling in the Ir substrate,
plays a crucial role to stabilize skyrmions.
The enhancement of the DM or the spin-orbit interactions at interfaces and, thereby, 
the emergence of atomic-scale skyrmions are actively studied recently~\cite{Chen2013,Kim2013,Rohart2013,Li2014,Dupe2014,Sonntag2014,Banerjee2014,Bergmann2014,Brede2014,Di2015,Bergmann2015,Schlenhoff2015, Romming2015}.
Frustrated spin-exchange interactions~\cite{Okubo2012} and nanopatterned magnetic thin film~\cite{Sun2013,Sapozhnikov2015}
are also predicted to accommodate a stable SkX.

Mathematically, a magnetic skyrmion is a two-dimensional (2D) configuration
of a three-dimensional (3D) unit vector field (whose manifold is isomorphic to the two-sphere $S^2$),
which is classified by the second homotopy group $\pi_2(S^2)$.
Hence, the skyrmions observed in bulk CMs are cylindrical configurations of skyrmionic structures,
which are not stabilized in the ground state but appear in a small region in the $B$-$T$ phase diagram just below the ferromagnetic phase transition temperature~\cite{Muhlbauer2009}.
It was predicted that the skyrmion crystal state can be the ground state of the two-dimensional CMs~\cite{Yi2009,Han2010,Li2011}.
Correspondingly, the region of the SkX phase is greatly enhanced down to $T=0$ in thin films~\cite{Yu2010,Yu2011}.
The SkX phase is further stabilized
in epitaxial thin films due to the magnetic anisotropy~\cite{Huang2012,Li2013}.
3D magnetization configurations are theoretically investigated in bulk and thin films of CMs,
where the multi-$\bm q$ configuration in the 3D reciprocal space and twisting of the skyrmionic structure
are discussed~\cite{Binz2006PRL,Binz2006PRB,Borisov2010,Park2011,Bogdanov2013,Bogdanov2014a,Bogdanov2014b}.
The experimental result in Ref.~\cite{Kanazawa2012} suggests that one of the 3D configurations may be realized in the bulk MnGe.
Appearance of a magentic monopole in merging dynamics of two skyrmions is also discussed in Ref.~\cite{Milde2013}.

In this paper, we theoretically show that by creating a hybrid system of a CM and a ferromagnet (FM),
various 3D magnetization configurations appear in the ground state.
Here, we consider a thin CM and assume that the magnetization is uniform along the $z$ direction (the direction perpendicular to the CM/FM interface) within the CM.
As in the case of a 2D CM, 
helical or skyrmionic structures appear in the CM 
at a low magnetic field.
However, the presence of the FM influences the CM as indicated by the reduced critical magnetic field below which skyrmions appear. 
The non-uniform structures appearing in the CM penetrate into the adjacent FM at a short distance from the interface.
Hence, a simple helix and a skyrmion-cylinder crystal (SCyX) appear when the FM is thin, which are the uniform configurations along the $z$ direction
and essentially the same as the spin helix and SkX in a 2D CM.
As the thickness of the FM increases, these structures become unstable and deform inside the FM.
When a spin helix arises in the CM, the helical structure is unwound in the FM by three-dimensionally rotating the magnetization vector,
forming a sideways half-cylinder skyrmion per helical period, which we call a sideways-skyrmion array (SSA).
On the other hand, for the case when a SkX appears in the CM, the skyrmionic structure shrinks as one goes deep inside the FM, ending up with a singularity of the magnetization, that is, a monopole.
We call a crystalline structure of such configurations a skyrmion-cone crystal (SCoX).
We also consider the case when the FM is sandwiched between two CMs with opposite signs of the DM interactions.
In this case, the helical or skyrmionic structures with opposite helicities appearing in the two CMs are continuously connected 
by twisting the magnetization vector in the FM along the $z$ direction, resulting in a twisted helix (TH) or a twisted-skyrmion crystal (TSX).

By minimizing the total energy for each configuration mentioned above,
we obtain the ground-state phase diagrams of the CM/FM and CM/FM/CM hybrid systems.
Here, we consider within a framework of the variational method where
the skyrmion radius, the helical pitch, and the penetration depth of the non-uniform structure are used as variational parameters.
We also calculate the emergent magnetic field that effectively acts on conduction electrons in the strong coupling limit,
and find that the emergent magnetic field takes nontrivial configurations due to the $z$ dependence of the helical and skyrmionic structures:
For example, the TH induces a staggered emergent magnetic field, whereas the emergent magnetic field for the SCoX points to or from the monopole and diverges at the monopole.

The rest of this paper is organized as follows.
In Sec.~\ref{sec:2D}, we review the phase diagram of a 2D CM with defining the characteristic energy and length scales.
The variational method used in the subsequent sections is also introduced in this section.
In Secs.~\ref{sec:CF} and \ref{sec:CFC}, we discuss the ground-state magnetization configurations, together with the emergent magnetic field,
at CM/FM heterostructures and CM/FM/CM hybrid structures, respectively,
by comparing the energies of possible magnetization configurations.
Discussions and conclusions are given in Sec.~\ref{sec:discussion}.

\section{Ground-state Phase Diagram of a 2D CM}
\label{sec:2D}

We first review the ground-state phase diagram of a 2D CM.
We consider a thin film of CM with thickness $a$ and 
assume that the magnetization along the thickness direction (the $z$ direction) is uniform.
We also assume that the emergent structure in the $xy$ plane is much larger than the atomic scale so that 
the energy of the system is described using the continuum model as~\cite{Bak1980}
\begin{align}
 E=a\iint dx dy \left[\frac{J}{2}(\nabla \nn)^2 - D \nn\cdot (\nabla\times \nn)+B(1-n_z)\right],
\label{eq:E2D}
\end{align}
where $\nn(x,y)$ is a unit vector describing the direction of the local magnetization,
$J(>0)$ and $D$ are the strengths of the spin-exchange and DM interactions, respectively,
and $B$ is the external magnetic field applied in the $z$ direction.
Here, the origin of the energy is chosen so that the ferromagnetic state, $\nn = \hat{z}$, has zero energy.
The DM interaction favors a non-uniform magnetization configurations ($\nabla\times \nn || \nn$), whereas 
the spin-exchange and Zeeman terms are minimized for a uniform magnetization aligned in the $z$ direction.
The competition between these terms results in the nontrivial magnetic structure of SkX.

\subsection{Spin Helix}
In low magnetic fields, a spin helix appears as a ground state,
where the magnetization $\nn$ winds lying in the perpendicular plane to the modulation vector $\kk$
so that it satisfies $\nabla\times \nn || \nn$.
Taking $\kk=k\hat{x}$, the magnetization profile is given by
\begin{align}
 \nn(x,y) = 
\begin{pmatrix}
0 \\ -\sin kx \\ -\cos kx
\end{pmatrix},
\label{eq:n_hel2D}
\end{align}
whose energy is obtained as
\begin{align}
 E_{\rm hel2D}(k)=aL^2\left(\frac{J}{2}k^2-Dk+B\right),
\label{eq:E_hel2D_k}
\end{align}
where $L$ is the system size in the $x$ and $y$ directions.
Here, the phase of the helix in Eq.~\eqref{eq:n_hel2D} is chosen so as to satisfy $\nn(0,0)=-\hat{z}$ for the sake of convenience in the latter discussions.
By minimizing Eq~\eqref{eq:E_hel2D_k} with respect to $k$,
we obtain the optimized wave number and energy as
\begin{align}
 k_{\rm hel2D}&=\frac{D}{J},\\
 E_{\rm hel2D}^0&=aL^2\left( - \frac{D^2}{2J}+B\right).
\label{eq:E_hel2D}
\end{align}

\subsection{Skyrmion Crystal}
\begin{figure}[tbp]
\includegraphics[width=0.99\linewidth]{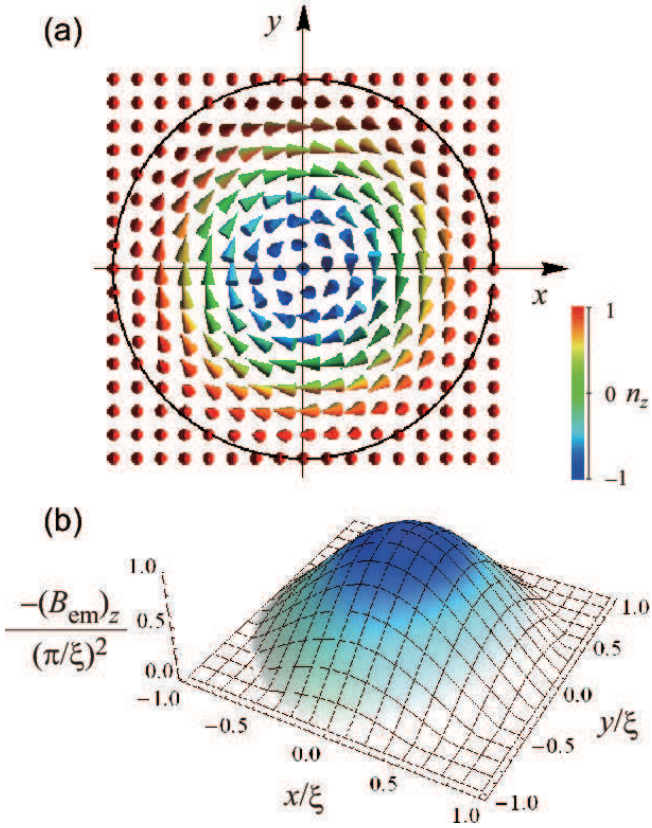}
\caption{(Color online) 
(a) Magnetization profile of a single skyrmion projected onto the $xy$ plane, where the colors on the arrows indicate the values of $n_z$.
The profile is given by Eq.~\eqref{eq:n_sk} with $\phi=-\pi/2$ and $\theta(r)=\pi(1-r/\xi)$,
which has the nonzero skyrmion charge, $\frac{1}{4\pi}\iint dx dy \nn\cdot(\partial_x\nn\times \partial_y\nn)=-1$.
The solid circle shows the radius of the skyrmion, $r=\xi$.
(b) Emergent magnetic field for the magnetization configuration shown in (a).
}
\label{fig:2D_Sk1}
\end{figure}

Since the helical structure has no net magnetization, it cannot survive under a large $B$ and instead, skyrmions appear.
We first consider an isolated skyrmion.
The magnetization profile around a charge $-1$ skyrmion is described in the 2D polar coordinates $(r,\varphi)$ as
\begin{align}
 \nn(r,\varphi) = 
\begin{pmatrix}
\sin\theta(r)\cos(\varphi+\phi)\\ \sin\theta(r)\sin(\varphi+\phi)\\ \cos\theta(r) 
\end{pmatrix},
\label{eq:n_sk}
\end{align}
where $\phi$ is a constant independent of $r$ and $\varphi$,
and $\theta(r)$ is a monotonically decreasing function that satisfies $\theta(0)=\pi$ and $\theta(r)=0$ for $r\ge \xi$
with $\xi$ being the radius of the skyrmion.
The magnetization profile described by Eq.~\eqref{eq:n_sk} is shown in Fig.~\ref{fig:2D_Sk1}(a).
Here, we introduce a function $\theta_0(\br)\ (0\le \br\le 1)$ such that $\theta(r) = \theta_0(r/\xi)$ for $0\le r \le \xi$;
$\theta_0(\br)$ monotonically decreases and satisfies $\theta_0(0)=\pi$ and $\theta_0(1)=0$.
In the following discussion, we fix $\theta_0(\br)=\pi(1-\rho)$ and take the skyrmion radius $\xi$ as a variational parameter.

Substituting Eq.~\eqref{eq:n_sk} in Eq.~\eqref{eq:E2D}, the energy for a single skyrmion is given by
\begin{align}
 E_{\rm Sk1}(\xi)&= a\left(\frac{\AJ  J}{2} + \AD  D \xi \sin\phi + \AB  B \xi^2\right),
\label{eq:E_Sk1_xi}
\end{align}
where 
\begin{align}
 \AJ  &\equiv 2\pi \int_0^1 \br d \br \left[\left(\frac{d\theta_0}{d\br}\right)^2 +  \frac{\sin^2\theta_0(\br)}{\br^2}\right], \\
 \AD  &\equiv -2\pi \int_0^1 \br d \br \left[\frac{d\theta_0}{d\br}+\frac{\sin2\theta_0(\br)}{2\br}\right], \\
 \AB  &\equiv 2\pi \int_0^1 \br d\br [1-\cos\theta_0(\br)].
\label{eq:def_AB}
\end{align}
Using $\theta_0(\br)=\pi(1- \br)$, these coefficients are given by
$\AJ =\pi[\pi^2+\gamma-{\rm Ci}(2\pi)+\log(2\pi)]\sim \pi(\pi^2+2.44), \AD = \pi^2$, and $\AB =\pi(1-4/\pi^2)$,
with $\gamma$ being the Euler's constant and ${\rm Ci}(z)$ the cosine integral function.
From the second term in the right-hand side of Eq.~\eqref{eq:E_Sk1_xi}, we find
$\phi=-\pi/2$ ($\phi=\pi/2$) for $D>0$ ($D<0$).
In the following discussions, we choose $D>0$ without loss of generality.

When the energy for a single skyrmion becomes negative, the system tends to create more skyrmions.
Hence, skyrmions in the ground state form a crystalline structure.
The total energy for the SkX is evaluated by multiplying the number of skyrmions $L^2/(\pi \xi^2)$ to Eq.~\eqref{eq:E_Sk1_xi}:
\begin{align}
 E_{\rm SkX}(\xi) = \frac{aL^2}{\pi}\left(\frac{\AJ  J}{2\xi^2} - \frac{\AD  D}{\xi} + \AB  B \right),
\label{eq:E_SkX_xi}
\end{align}
where we used $\phi=-\pi/2$.
Minimizing Eq.~\eqref{eq:E_SkX_xi} with respect to $\xi$,
the optimized skyrmion radius and the energy of the SkX are respectively given by
\begin{align}
 \xi_{\rm SkX} &= \frac{\AJ }{\AD }\frac{J}{D},\\
 E_{\rm SkX}^0  &= \frac{aL^2}{\pi}\left[-\frac{\AD^2 D^2}{2\AJ J}+ \AB  B  \right].
\label{eq:E_SkX}
\end{align}

For $\theta_0(\br)=\pi(1- \br)$, we obtain $\xisk\simeq 3.9 J/D$.
On the other hand, a SkX is described with a superposition of three helical spin textures
with the modulation vectors satisfying $|\bm k_{i=1,2,3}|=k_{\rm hel2D}$ and $\sum_{i=1,3}\bm k_i=\bm 0$~\cite{Muhlbauer2009},
which leads to a skyrmion radius (a half of the triangular lattice constant) $\xi=2\pi/(\sqrt{3}k_{\rm hel2D})\simeq 3.6 J/D$.
The small difference between these values suggests that the actual profile of $\theta_0(\br)$ does not so deviate from $\pi(1-\br)$.

\subsection{Phase Diagram}
\begin{figure}[tbp]
\includegraphics[width=0.99\linewidth]{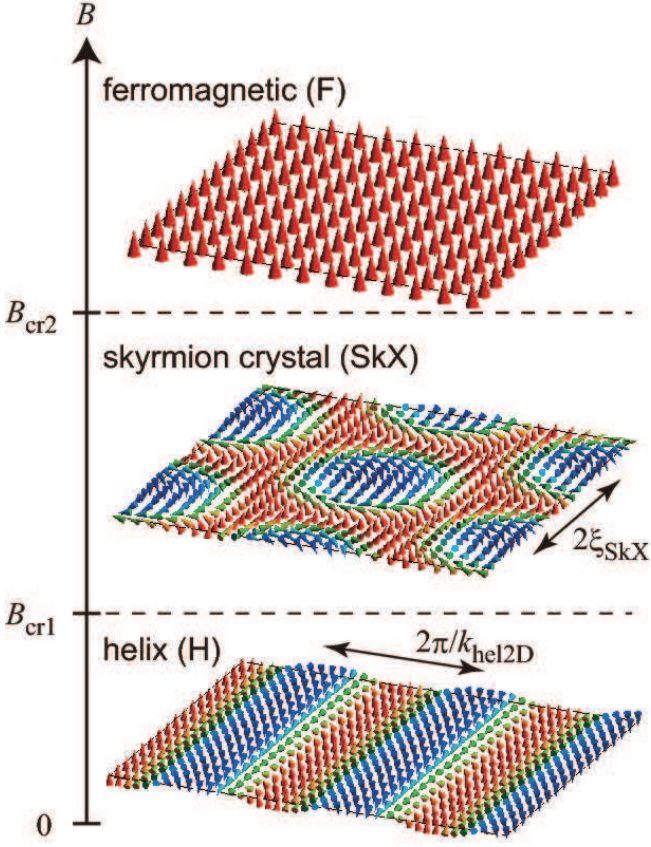}
\caption{(Color online) Schematic phase diagram of a two-dimensional chiral magnet and magnetization structure in each phase.}
\label{fig:PD_2D}
\end{figure}

By comparing the energies for the spin helix [Eq.~\eqref{eq:E_hel2D}], the SkX [Eq.~\eqref{eq:E_SkX}], and the ferromagnetic state [$E_{\rm ferro}=0$], 
the magnetic structure in the ground state changes as
\begin{align}
 0<B<B_{\rm cr1}&:\ \ \ \textrm{helix},\\
 B_{\rm cr1}<B<B_{\rm cr2}&:\ \ \ \textrm{SkX},\\
 B_{\rm cr2}<B&:\ \ \ \textrm{ferromagnet},
\end{align}
where we have assumed the system size is larger than the area of a skyrmion ($L^2>\pi\xi_{\rm SkX}^2$),
and the critical magnetic fields are defined as
\begin{align}
 B_{\rm cr1}&\equiv \frac{\pi-\AD ^2/\AJ }{\pi-\AB }\frac{D^2}{2J},\\
 B_{\rm cr2}&\equiv \frac{\AD ^2 D^2}{2\AB \AJ J }  \ \ \ (\,>B_{\rm cr1}).
\end{align}
The schematic phase diagram is shown in Fig.~\ref{fig:PD_2D}.
Although the actual profile of $\theta_0(\br)$ depends on $B$,
our variational function with a fixed $\theta_0(\br)$ can capture the ground-state property of the 2D CM.
In particular, the critical values obtained by using $\theta_0(\br)=\pi(1-\br)$ are $B_{\rm cr1}=0.24D^2/J$ and $B_{\rm cr2}=0.67D^2/J$,
which reasonably agree with the numerically obtained ones
$B_{\rm cr1}=0.23D^2/J$ and $B_{\rm cr2}=0.78D^2/J$~\cite{Mochizuki2012,Iwasaki2013a}.

We should remark here that the above discussion is valid only for a thin film as the SkX phase disappears from the ground-state phase diagram when $a \gg \xisk$.
This is because a conical structure, which is a spin helix along the $z$ direction with uniform longitudinal magnetization, has lower energy than the SkX.
In experiments, the SkX phase is observed in the ground state up to $a\sim 4\xisk$~\cite{Seki2012S}.

\subsection{Emergent Magnetic Field}
One of the striking effects of the appearance of the SkX is that it causes the emergent electromagnetic field, 
which then leads to the topological Hall effect and the electromagnetic induction~\cite{NagaosaTokura2013}.
Suppose that the conduction electron spin is coupled to, and forced to be parallel to, the localized magnetization.
In the strong coupling limit, the electrons behave as if there is an emergent electromagnetic field defined by
\begin{align}
 (B_{\rm em})_i &= \frac{1}{2}\sum_{j,k=x,y,z}\epsilon_{ijk}\nn\cdot\left(\partial_j\nn\times \partial_k\nn\right),
\label{eq:Bem_def}\\
 (E_{\rm em})_i &= \nn \cdot\left(\partial_i \nn \times \partial_t \nn\right),
\end{align}
where $\partial_i=\partial/\partial x_i$ and $\epsilon_{ijk}$ is the totally antisymmetric tensor in three dimensions.
In the static magnetization configuration, we have $\bm E_{\rm em}=\bm 0$, whereas $\bm B_{\rm em}$ is non-vanishing for non-coplanar configurations.
Indeed, we obtain $\bm B_{\rm em}=\bm 0$ for the spin helix [Eq.~\eqref{eq:n_hel2D}] and
\begin{align}
 \bm B_{\rm em} =  \frac{\sin\theta}{r}\frac{d\theta}{dr} \hat{z} =  -\left(\frac{\pi}{\xi}\right)^2 \frac{\sin(\pi r/\xi) }{\pi r/\xi} \hat{z},
\label{eq:Bem_Sk1}
\end{align}
for the skyrmionic configuration [Eq.~\eqref{eq:n_sk}],
where in the last equality in Eq.~\eqref{eq:Bem_Sk1} we used $\theta(r) = \pi(1-r/\xi)$. 
The distribution of the emergent magnetic field given by Eq.~\eqref{eq:Bem_Sk1} is shown in Fig.~\ref{fig:2D_Sk1}(b).

\subsection{Dimensionless Parameters}
In the following sections, we scale the length in units of $\xisk$ and the energy in units of $\AJ  J L^2/(2\pi\xisk)$.
The dimensionless variables are denoted with tilde, e.g.,
\begin{align}
 \txi &= \frac{\xi}{\xisk},\\
 \tk  &= \xisk k,\\
 \tE  &= \frac{2\pi \xisk E}{\AJ  JL^2}.
\end{align}
We also introduce a scaled magnetic field
\begin{align}
b\equiv \frac{B}{B_{\rm cr2}}.
\end{align}
Using these notations, Eqs.~\eqref{eq:E_hel2D_k} and \eqref{eq:E_SkX_xi} are respectively rewritten as
\begin{align}
 \tE_{\rm hel2D}(\tk)  &= \pi\ta\left(\frac{\tk^2}{\AJ }-\frac{2\tk}{\AD }+\frac{b}{\AB }\right), \label{eq:tE_hel2D_tk}\\
 \tE_{\rm SkX}(\txi) &= \ta\left(\frac{1}{\txi^2}-\frac{2}{\txi}+b\right),
\end{align}
and the scaled value for the critical field $B_{\rm cr1}$ is given by
\begin{align}
 b_1 \equiv \frac{B_{\rm cr1}}{B_{\rm cr2}} = \frac{\AJ \AB}{\AD^2}\frac{\pi-\AD^2/\AJ}{\pi-\AB}.
\label{eq:b1}
\end{align}

\section{CM/FM heterostructure}
\label{sec:CF}
\begin{figure}[tbp]
\includegraphics[width=0.9\linewidth]{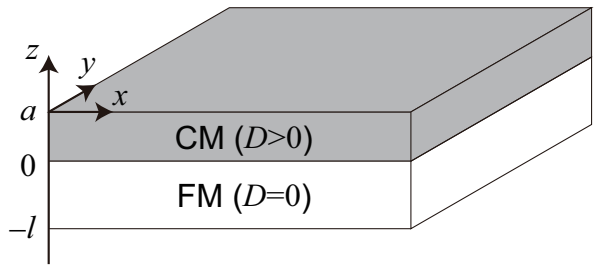}
\caption{Schematic of the CM/FM heterostructure,
where CM and FM denote chiral magnet and ferromagnet, respectively,
and $D$ is the coupling constant of the Dzyaloshinskii-Moriya interaction [See Eq.~\eqref{eq:Hamiltonian_CF}].
}
\label{fig:CF_sys}
\end{figure}

We consider a heterostructure of a CM with thickness $a$ on a FM with thickness $l$ (Fig.~\ref{fig:CF_sys}).
For simplicity, we assume that the ferromagnetic interaction is the same in the whole system.
The total energy of the system is given by
\begin{align}
 E=&\int_{-l}^a dz \iint dx dy  \,\left[\frac{J}{2}(\nabla \nn)^2 +B(1-n_z)\right] \non\\
&-D \int_0^a dz \iint dx dy\,  \nn\cdot (\nabla\times \nn).
\label{eq:Hamiltonian_CF}
\end{align}

\begin{figure*}[tbp]
\includegraphics[width=0.99\linewidth]{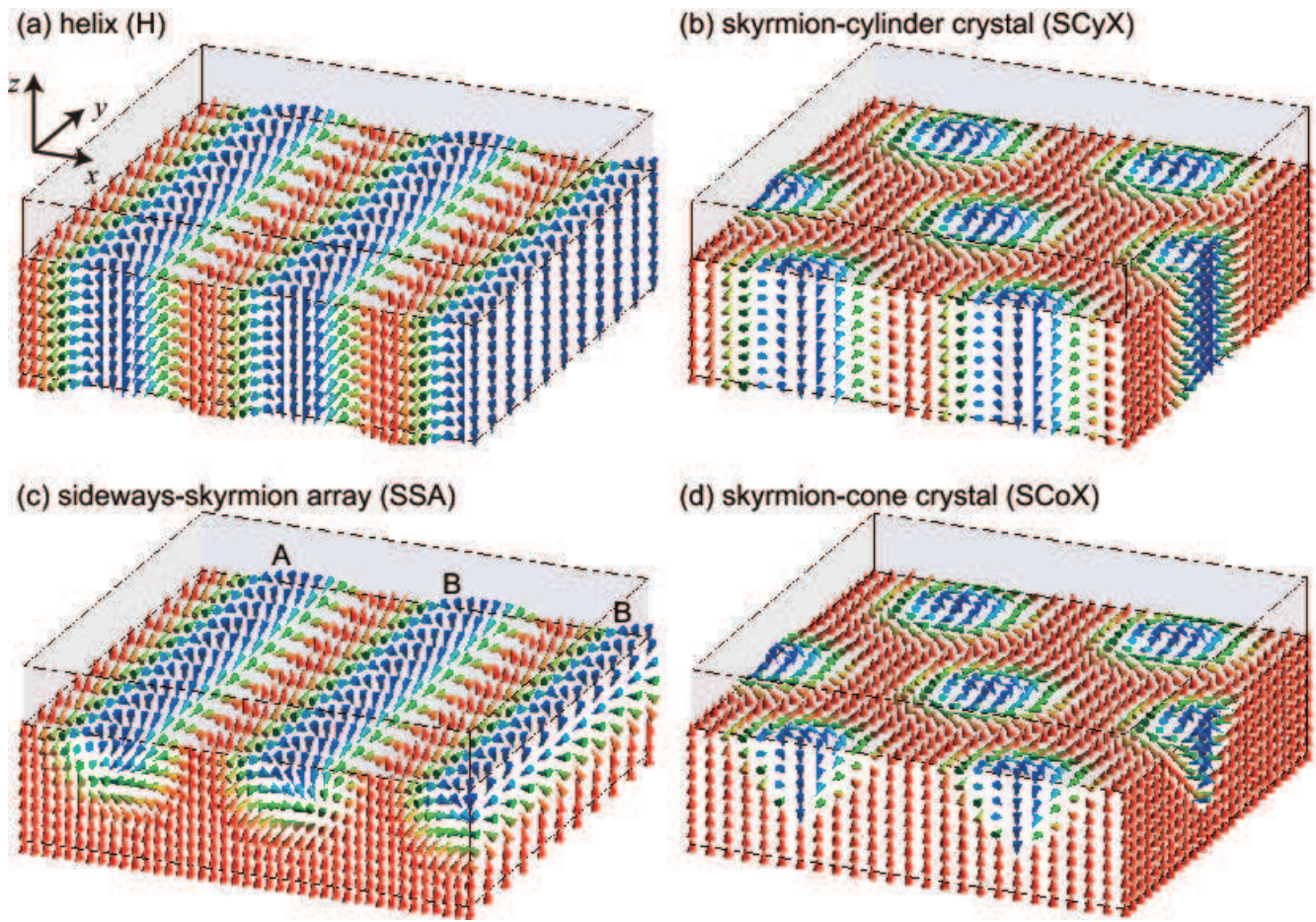}
\caption{(Color online) 
Possible three-dimensional magnetization structures in chiral magnet (CM)/ferromagnet (FM) heterostructures,
where the magnetization vectors in the FM are shown.
The magnetization in the CM, which is shown with a gray cuboid, is assumed to be uniform along the $z$ direction
and forms a spin helix [(a) and (c)] or a skyrmion crystal [(b) and (d)].
The magnetization configurations in (a) and (b) are independent of $z$,
whereas those in (c) and (d) deform as a function of $z$ and become uniform at a distance from the CM/FM interface.
In (c), the rows indicated by A and B have opposite whirling patterns in the $xz$ plane, 
whose magnetization profiles are given by $\bm n_+(x,y,z)$ and $\bm n_-(x,y,z)$ defined in Eq.~\eqref{eq:n_swsk}, respectively.
The whirling direction is spontaneously chosen for row by row.}
\label{fig:CF_n} 
\end{figure*}

\begin{figure*}[tbp]
\includegraphics[width=0.99\linewidth]{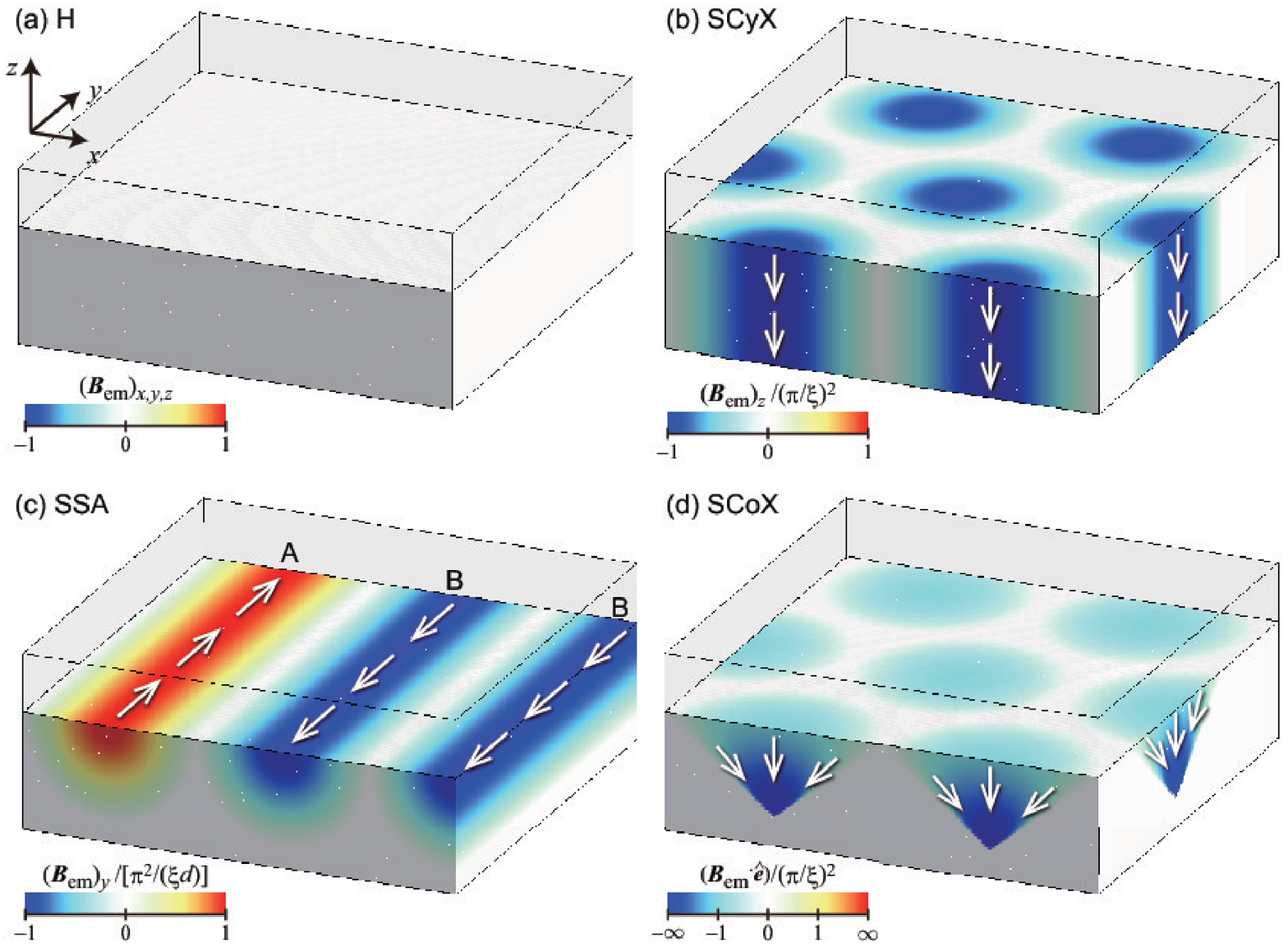}
\caption{(Color online) 
Emergent magnetic field $\bm B_{\rm em}$ for the magnetization configurations shown in Fig.~\ref{fig:CF_n},
where H, SCyX, SSA, SCoX stand for helix, skyrmion-cylinder crystal, sideways-skyrmion array, and skyrmion-cone crystal, respectively.
${\bm B}_{\rm em}$ for (b), (c), and (d) are given by Eqs.~\eqref{eq:Bem_Sk1}, \eqref{eq:Bem_SSA}, and \eqref{eq:Bem_SCoX}, respectively.
In each panel, the color scale for the emergent magnetic field is scaled by its maximum, 
where $\xi$ and $d$ are the skyrmion radius (or the half of the helical pitch) and the penetration depth of the non-uniform structure, respectively.
The direction of the emergent magnetic field is schematically shown with white arrows.
(a) There is no emergent magnetic field in any direction for the spin helix.
(b) For the SCyX, the emergent magnetic field in the $-z$ direction arises at around the center of the skyrmion.
(c) For the SSA, the emergent magnetic field arises in the $y$ direction and its sign depends on the whirling direction in the $xz$ plane. 
The rows indicated by A and B in (c) corresponds to those in Fig.~\ref{fig:CF_n}(c), which have opposite whirling patterns in the $xz$ plane, and hence, the direction of the emergent magnetic field is opposite.
(d) The emergent magnetic field for the SCoX is non-collinear and parallel to $\hat{\bm e}$, a unit vector along $\rr-\rr_{\rm m}$ with $\rr_{\rm m}$ being the position of the nearest monopole.
The amplitude of the emergent magnetic field shown in (d) diverges at the monopoles.
}
\label{fig:CF_B}
\end{figure*}

The 3D magnetization configurations that minimize Eq.~\eqref{eq:Hamiltonian_CF} and the resulting emergent magnetic fields
are summarized in Figs.~\ref{fig:CF_n} and \ref{fig:CF_B}, respectively.
Here, we assume that the magnetization profile in the CM is uniform along the $z$ direction.
As we saw in the previous section, there are two possible 2D configurations in the CM, i.e., the spin helix and the SkX.
For each configuration, there are two possible configurations in the adjacent FM:
When the thickness of the FM is small, the magnetization configuration in the CM uniformly penetrates into the FM [Fig.~\ref{fig:CF_n}(a) and \ref{fig:CF_n}(b)];
On the other hand, when the FM is thick enough, the magnetization configuration is deformed in the FM and disappears at a finite depth from the CM/FM interface [Fig.~\ref{fig:CF_n}(c) and \ref{fig:CF_n}(d)].
We calculate the energy and the emergent magnetic field for the each configuration in Secs.~\ref{sec:CM/FM_helix_uniform}--\ref{sec:CM/FM_SkX_z}
and discuss the phase diagram in Sec.~\ref{sec:CM/FM_PD}.

\subsection{Spin Helix}
\label{sec:CM/FM_helix_uniform}
We first consider the case when a spin helix appearing in the CM uniformly penetrates into the FM.
When the magnetization vector is given by Eq.~\eqref{eq:n_hel2D} for all $-l\le z\le a$,
the total energy is given by
\begin{align}
 E_{\rm hel3D}(k)&=L^2\left[(a+l)\frac{J}{2}k^2 - aDk + (a+l)B\right],
\label{eq:E_hel3D_k}
\end{align}
or, equivalently, 
\begin{align}
 \tE_{\rm hel3D}(\tk) &= \pi(\ta+\tl)\left(\frac{\tk^2}{\AJ}-\frac{\ta}{\ta+\tl}\frac{2\tk}{\AD}+\frac{b}{\AB}\right).
\end{align}
By minimizing $\tE_{\rm hel3D}(\tk)$ with respect to $\tk$, we obtain the optimized wave number and energy as
\begin{align}
 \tk_{\rm hel3D}   &= \frac{\ta}{\ta+\tl}\frac{\AJ}{\AD} = \frac{\ta}{\ta+\tl}\tk_{\rm hel2D},
\label{eq:k_hel3D}\\
 \tE_{\rm hel3D}^0 &= \pi (\ta+\tl)\left[\frac{b}{\AB}-\left(\frac{\ta}{\ta+\tl}\right)^2\frac{\AJ}{\AD^2}\right].
\label{eq:E_hel3D}
\end{align}
As one can see from the comparison between Eqs.~\eqref{eq:E_hel2D_k} and \eqref{eq:E_hel3D_k}, the effective DM interaction relative to the ferromagnetic and Zeeman interactions
in the CM/FM heterostructure is decreased by a factor $\ta/(\ta+\tl)$,
resulting in the reduction of the optimized wave number as shown in Eq.~\eqref{eq:k_hel3D}.
As in the case of the 2D CM, there is no emergent magnetic field for the spin helix [Fig.~\ref{fig:CF_B}(a)].

\subsection{Skyrmion-Cylinder Crystal}
\label{sec:CM/FM_SkX_uniform}
Similar to Sec.~\ref{sec:CM/FM_helix_uniform}, when a skyrmion appears in the CM and uniformly penetrates into the FM,
the magnetization vector is given by Eq.~\eqref{eq:n_sk} for all $-l\le z\le a$.
A possible candidate for the ground state is the crystalline structure of such configurations, i.e., the SCyX.
The energy for the SCyX is calculated in the same manner as Eq.~\eqref{eq:E_SkX_xi}.
In the present case, the system size along the $z$ direction is elongated by a factor $(a+l)/a$ and the effective DM interaction relative to the other interactions is reduced by a factor $a/(a+l)$.
As a result, we obtain 
\begin{align}
 E_{\rm SCyX}(\xi)&= \frac{(a+l)L^2}{\pi}\left(\frac{\AJ  J}{2\xi^2} - \frac{a}{a+l}\frac{\AD D}{\xi} + \AB  B\right),
\end{align}
which reduces to
\begin{align}
\tE_{\rm SCyX}(\txi)
&= (\ta+\tl )\left(\frac{1}{\txi^2} - \frac{\ta}{\ta+\tl }\frac{2}{\txi} + b\right).
\label{eq:ESCyX_xi}
\end{align}
By minimizing $\tE_{\rm SCyX}(\txi)$ with respect to $\txi$, we obtain the optimized skyrmion radius and energy as
\begin{align}
 \txi_{\rm SCyX} &= \frac{\ta+\tl}{\ta},\\
 \tE_{\rm SCyX}^0 &= (\ta+\tl )\left[b-\left(\frac{\ta}{\ta+\tl}\right)^2\right].
\label{eq:E_SCyX}
\end{align}

The emergent magnetic field for a single skyrmion cylinder is the same as that for a skyrmion in a 2D CM and given by Eq.~\eqref{eq:Bem_Sk1}. 
Figure~\ref{fig:CF_B}(b) shows the configurations of $\bm B_{\rm em}$ for the SCyX.

\subsection{Sideways-Skyrmion Array}
\label{sec:CM/FM_helix_z}
\begin{figure}[tbp]
\includegraphics[width=0.7\linewidth]{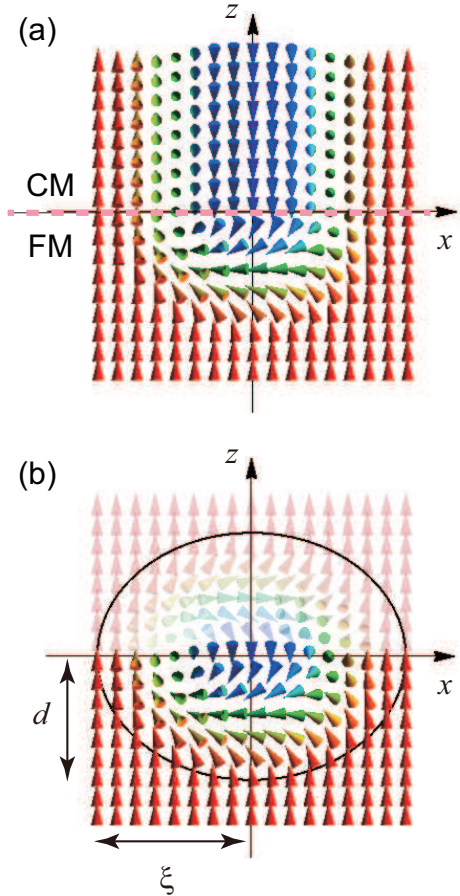}
\caption{(Color online) 
(a) Magnetization profile of a sideways half-cylinder skyrmion given by $\bm n_+(x,y,z)$ in Eq.~\eqref{eq:n_swsk}.
Shown are the magnetization vectors projected onto the $xz$ plane.
(b) The same as (a) but the magnetization configuration at $z>0$ is replaced
so as to form a full skyrmion on the $xz$ plane.
Namely, $\nn$ in $z>0$ is defined as $n_x(x,y,z)=-n_x(x,y,-z)$ and $n_{y,z}(x,y,z)=n_{y,z}(x,y,-z)$.
The solid circle shows the region of the skyrmion, $(x/\xi)^2+(z/d)^2=1$.
The skyrmion charge for the configuration in (b)
is given by $\frac{1}{4\pi}\iint dxdz \nn\cdot(\partial_z\nn\times \partial_x \nn)=1$.
}
\label{fig:CF_SSA2}
\end{figure}

When the FM is thick enough, it is not energetically favorable to keep the non-uniform magnetization configuration in the whole FM.
The non-uniform configuration appearing in the CM penetrates only into a finite depth of the FM.
For the case when a spin helix appears in the CM, the helical structure is unwound by three-dimensionally rotating the magnetization vector as shown in Fig.~\ref{fig:CF_n}(c).
This is nothing but a one-dimensional array of sideways half-cylinder skyrmions.
We pick up one of the sideways skyrmion at $y=0$ and consider the following ansatz:
\begin{align}
&\bm n_\pm(x,y,z) \non\\
&=\left\{
\begin{array}{ll}
\begin{pmatrix}
0 \\ -{\rm sgn}(x)\sin\vartheta(|x|) \\ \cos\vartheta(|x|)
\end{pmatrix} & (0\le z \le a), 
 \\[7mm]
\begin{pmatrix}
\sin\vartheta(\eta) \cos(\pm\chi+\phi) \\ \sin\vartheta(\eta) \sin(\pm\chi+\phi) \\ \cos\vartheta(\eta)
\end{pmatrix} & (-l \le z \le 0),
\end{array}\right.
\label{eq:n_swsk}
\end{align}
where $\eta=\sqrt{x^2+(\xi z/d)^2}$, $\chi=\arg(x+i\xi z/d)$, $\phi=-\pi/2$, and $\vartheta(\eta)=\theta_0(\eta/\xi)$.
Here, we consider an elliptically deformed skyrmion and $\xi$ and $d$ are the radius in the $x$ and $z$ directions, respectively.
We choose $\theta_0(\br) = \pi (1-\br)$. Then, Eq.~\eqref{eq:n_swsk} at $z\ge 0$ coincides with Eq.~\eqref{eq:n_hel2D} with $k=\pi/\xi$.
In Fig.~\ref{fig:CF_SSA2}, we plot the magnetization vector field given by $\bm n_+(x,y,z)$ in Eq.~\eqref{eq:n_swsk},
from which one can see that the helical structure is continuously deformed to a uniform one
and that this is indeed a half of a skyrmion.
Note that though similar configurations have been considered in Refs.~\cite{Yokouchi2014,Yokouchi2015}, the present configuration is distinct from them
as the axis of the skyrmion and, thereby, the emergent magnetic field are perpendicular to the external magnetic field in the SSA,
whereas in Refs.~\cite{Yokouchi2014,Yokouchi2015} the axis of the skyrmions are parallel to the external magnetic field.

The topological charge for a sideways skyrmion is defined by the integral of the skyrmion density in the FM region and calculated as
\begin{align}
\frac{1}{4\pi}\int_{-d}^0 dz\int_{-\xi}^\xi dx \nn_\pm\cdot(\partial_z\nn_\pm\times \partial_x \nn_\pm)=\pm \frac{1}{2}. 
\end{align}
As we shall see below, within the framework of the variational method,
the energies for the configurations $\bm n_+$ and $\bm n_-$ are degenerate.
Hence, which configuration appears is spontaneously determined for row by row [see Fig.~\ref{fig:CF_n}(c)].

By substituting Eq.~\eqref{eq:n_swsk} into Eq.~\eqref{eq:Hamiltonian_CF}, 
the energy for a sideways skyrmion is given by
\begin{align}
E_{\rm SS1}(d,\xi) 
=& 2aL\xi\left[\frac{J}{2}\left(\frac{\pi}{\xi}\right)^2-D\left(\frac{\pi}{\xi}\right)+B\right]\non\\
&+L\left[\frac{\AJ J}{8}\left(\frac{d}{\xi}+\frac{\xi}{d}\right)+\frac{\AB B}{2}\xi d\right].
\label{eq:E_SS1_xi}
\end{align}
Here, the first term in the right-hand side of Eq.~\eqref{eq:E_SS1_xi} is the energy of the CM, which is given by Eq.~\eqref{eq:E_hel2D_k} with replacing the system size $L^2$ to $2\xi L$,
whereas the second term comes from the FM.
The total energy for the SSA is obtained by multiplying the number of half-cylinder skyrmions $L/(2\xi)$
and its dimensionless value is given by
\begin{align}
\tE_{\rm SSA}(\td,\txi) =& \pi\bigg(
 \frac{\ta\pi^2}{\AJ \txi^2}-\frac{2\ta\pi}{\AD \txi}+\frac{\ta b}{\AB}
+\frac{\td}{8\txi}+\frac{\txi}{8\td}+\frac{\td b}{4}
\bigg).
\label{eq:E_SSA_dxi}
\end{align}
Equation~\eqref{eq:E_SSA_dxi} has a minimum with respect to $\td$ at
\begin{align}
 \td_{\rm SSA} = \frac{\txi}{\sqrt{1+2b\txi^2}},
\label{eq:d_SSA}
\end{align}
and the total energy as a function of $\txi$ is given by
\begin{align}
&\tE_{\rm SSA}(\td_{\rm SSA},\txi) \non\\
&= \pi\left(
 \frac{\ta\pi^2}{\AJ \txi^2}-\frac{2\ta\pi}{\AD \txi}+\frac{\ta b}{\AB}
+\frac{\sqrt{1+2b\txi^2}}{4\txi}
\right).
\label{eq:E_SSA_xi} 
\end{align}
Equation~\eqref{eq:d_SSA} shows that $\td$ is in the same order as $\txi$ and decreases as $b$ increases,
which means that the sideways skyrmions are compressed to the interface so as to reduce the Zeeman energy.
In order to compair with other comfigurations, we numerically minimize Eq.~\eqref{eq:E_SSA_xi} with respect to $\txi$
and obtain the energy of the SSA.

Since a sideways skyrmion is a skyrmion in the $xz$ plane, it induces the emergent magnetic field in the $y$ direction.
By substituting Eq.~\eqref{eq:n_swsk} in Eq.~\eqref{eq:Bem_def}, we obtain 
\begin{align}
 {\bm B}_{{\rm em},\pm} &= \bm n_{\pm}\cdot \left(\partial_z\bm n_\pm \times \partial_x \bm n_\pm\right)\non\\
&= \pm \frac{\pi^2}{d\xi} \frac{\sin \pi\eta/\xi}{\pi\eta/\xi}\hat{y},
\label{eq:Bem_SSA}
\end{align}
where $\eta=\sqrt{x^2+(\xi z/d)^2}$.
The emergent magnetic field for the magnetization configuration in Fig.~\ref{fig:CF_n}(c) is shown in Fig.~\ref{fig:CF_B}(c),
where the rows indicated by A and B in Figs.~\ref{fig:CF_n}(c) and \ref{fig:CF_B}(c) correspond to each other.

\subsection{Skyrmion-Cone Crystal}
\label{sec:CM/FM_SkX_z}
\begin{figure}[tbp]
\includegraphics[width=0.9\linewidth]{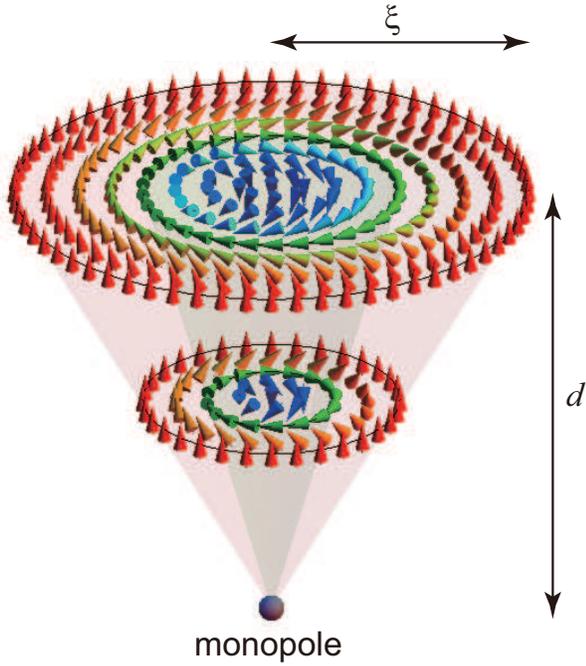}
\caption{(Color online) 
Magnetization profile in a skyrmion cone. The skyrmion radius shrinks as one goes inside the FM.
At the top of the cone, a monopole emerges.}
\label{fig:CF_SCoX2}
\end{figure}

Similar to Sec.~\ref{sec:CM/FM_helix_z}, when a SkX appears in the CM and the adjacent FM is thick enough,
the skyrmionic structure cannot penetrate into the whole FM [Fig.~\ref{fig:CF_n}(d)].
In Fig.~\ref{fig:CF_SCoX2}, we show the 3D structure developed below a skyrmion.
We call the structure shown in Fig.~\ref{fig:CF_SCoX2} a skyrmion cone.
When we see the 2D structure of the skyrmion cone perpendicular to the $z$ axis,
the skyrmionic structure shrinks as one goes deep inside the FM and eventually disappears at a finite depth $d$.
Note that because a skyrmion is a topologically nontrivial structure, 
it cannot disappear under a continuous deformation.
Hence, the skyrmionic configuration ends up with a defect of the magnetization, that is, a monopole.

To give a concrete profile of the magnetization,
we consider a skyrmion with radius $\xi$ in the region of $0 \le z \le a$, whose magnetization vector is given by Eq.~\eqref{eq:n_sk},
and assume that the skyrmion shrinks as a function of $z$ and disappears at $z=-d<0$.
The magnetization profile for $-d<z<0$ is given by Eq.~\eqref{eq:n_sk} with replacing $\theta(r)$ with the following $z$-dependent function:
\begin{align}
 \theta(r,z) = \theta_0\left(\frac{r}{\xi f(|z|/d)}\right),
\label{eq:theta_rz}
\end{align}
where $f(\zeta)$ is a monotonically decreasing function satisfying $f(0)=1$ and $f(1)=0$, and $\xi f(|z|/d)$ describes the skyrmion radius at depth $|z|$.
Substituting Eqs.~\eqref{eq:n_sk} and \eqref{eq:theta_rz} in Eq.~\eqref{eq:Hamiltonian_CF}, the total energy for a skyrmion cone is given by
\begin{align}
 E_{\rm SCo1}(d,\xi) =&  a\left(\frac{\AJ J}{2}-\AD D\xi + \AB B\xi^2\right)\non\\
&+\frac{d \AJ J}{2}+\frac{\xi^2\BJ J}{2d}+d\xi^2\BB B,
\label{eq:E_SCo1_dxi}
\end{align}
where the first and second lines of the right-hand side of Eq.~\eqref{eq:E_SCo1_dxi} correspond to the energies for the CM and the FM, respectively, 
and we have defined
\begin{align}
\BJ &\equiv 2\pi \int_0^1 d\zeta \left(\frac{df}{d\zeta}\right)^2 \int_0^1 \left(\frac{d\theta_0}{d\br}\right)^2\br^3d\br,\\
\BB &\equiv \AB  \int_0^1 d\zeta f^2(\zeta).
\end{align}
Here, we approximate $f(\zeta)=1-\zeta$ and $\theta_0(\br)=\pi (1-\br)$.
Then, the above coefficients are given by $\BJ =\pi^3/2$ and $\BB =\pi/3(1-4/\pi^2)$.

The total energy for a SCoX is obtained by multiplying the number of the skyrmion cones $L^2/(\pi\xi^2)$
to Eq.~\eqref{eq:E_SCo1_dxi}.
The dimensionless value is given by
\begin{align}
\tE_{\rm SCoX} (\td,\txi)= \frac{\ta}{\txi^2}-\frac{2\ta}{\txi}+\ta b + \frac{\td}{\txi^2}+\frac{\beta_J}{\td}+\beta_B\td b,
\label{eq:E_SCoX_dxi}
\end{align}
where  $\beta_{J}\equiv\BJ/\AJ\simeq 0.40$ and
$\beta_{B}\equiv \BB/\AB=1/3$.
Equation~\eqref{eq:E_SCoX_dxi} has a minimum with respect to $\td$ at 
\begin{align}
 \td_{\rm SCoX} 
=\frac{\sqrt{\beta_J} \txi}{\sqrt{1 + \beta_B b \txi^2}}.
\label{eq:d_SCoX}
\end{align}
Similar to Eq.~\eqref{eq:d_SSA}, $\td_{\rm SCoX}$ is in the same order as $\txi$ and decreases as $b$ increases,
which means that the penetration depth of skyrmions becomes smaller for larger $b$.
At $\td=\td_{\rm SCoX}$, the total energy is given by
\begin{align}
\tE_{\rm SCoX}(\txi)
& = \frac{\ta}{\txi^2} - \frac{2\ta}{\txi}  + \ta b + \frac{2\sqrt{\beta_J(1+\beta_B b\txi^2)}}{\txi}.
\label{eq:E_SCoX_xi}
\end{align}
In order to compair with other configurations, we numerically minimize Eq.~\eqref{eq:E_SCoX_xi} with respect to $\txi$ and obtain the energy of the SCoX.

Taking into account the $z$ dependence of $\theta$ and substituting Eq.~\eqref{eq:n_sk} in Eq.~\eqref{eq:Bem_def}, the emergent magnetic field is calculated as
\begin{align}
 {\bm B}_{\rm em} 
&=-\left[\frac{\pi}{\xi(z)}\right]^2 \frac{\sin[\pi r/\xi(z)]}{\pi r/\xi(z)} \frac{\rr - \rr_{\rm m}}{z+d},
\label{eq:Bem_SCoX}
\end{align}
where $\xi(z)\equiv \xi f(|z|/d)=\xi(1+z/d)$ is the $z$-dependent skyrmion radius, 
$\rr=(x,y,z)$, and $\rr_{\rm m}=(0,0,-d)$ is the position of the monopole.
The emergent magnetic field is non-vanishing inside the skyrmion cone.
It points to the monopole, and the amplitude diverges at the monopole.
The configuration of ${\bm B}_{\rm em}$ for the SCoX shown in Fig.~\ref{fig:CF_n}(d) is depicted in Fig.~\ref{fig:CF_B}(d).

\subsection{Phase Diagram}
\label{sec:CM/FM_PD}
\begin{figure}[tbp]
\includegraphics[width=0.95\linewidth]{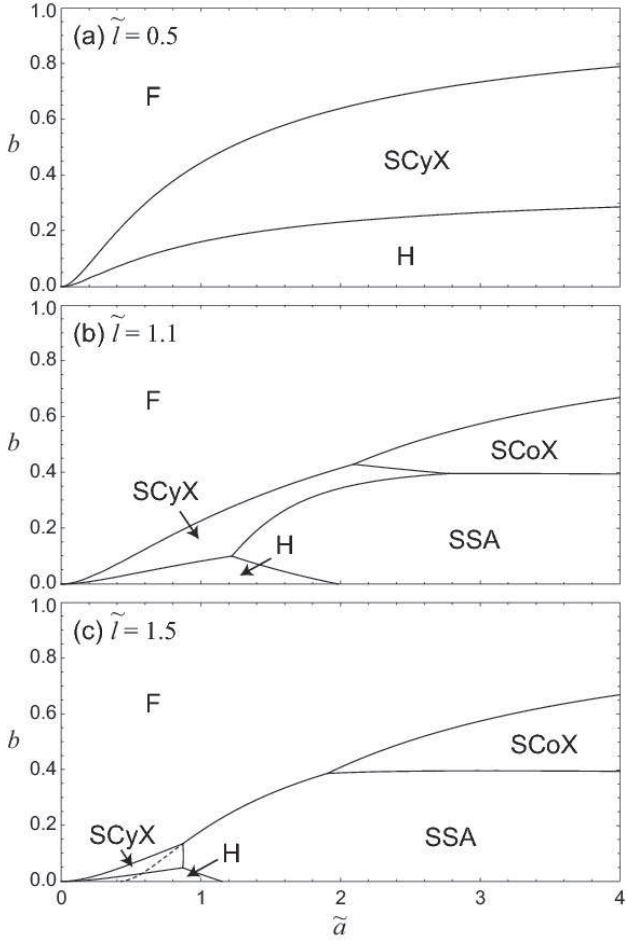}
\caption{Phase diagram of a CM/FM heterostructure for (a) $\tl=0.5$, (b) $\tl=1.1$, and (c) $\tl=1.5$.
Here, F, SCyX, H, SCoX, and SSA stand for ferromagnetic, skyrmion-cylinder crystal, helix, skyrmion-cone crystal, and sideways-skyrmion array phases, respectively.
The F--SCyX and SCyX--H phase boundaries are given by Eqs.~\eqref{eq:b_F-SCyX} and \eqref{eq:b_SCyX-H}, respectively.
The other phase boundaries are numerically calculated.
The dashed curve in (c) indicates the F--SSA phase boundary for $\tl\gg1$.
}
\label{fig:PD_CMFM}
\end{figure}

By comparing the energy for each configuration, the phase diagram of the CM/FM heterostructure
is obtained in the $(\ta, b)$ space as shown in Fig.~\ref{fig:PD_CMFM}.
Here, we calculate for (a) $\tl=0.5$, (b) $\tl=1.1$, and (c) $\tl=1.5$. 
As expected, for a small $\tl$ [Fig.~\ref{fig:PD_CMFM}(a)], only the helix (H) and SCyX phases appear.
In this case, the phase boundaries are analytically obtained by comparing the energies in Eqs.~\eqref{eq:E_hel3D} and \eqref{eq:E_SCyX} and $E_{\rm F}=0$,
and given by
\begin{align}
 b_\textrm{F-SCyX}&=\left(\frac{\ta}{\ta+\tl}\right)^2,
\label{eq:b_F-SCyX}\\
 b_\textrm{SCyX-H}&=\left(\frac{\ta}{\ta+\tl}\right)^2 b_1,
\label{eq:b_SCyX-H}
\end{align}
where $b_1$ is the critical magnetic field at $\tl=0$ and defined in Eq.~\eqref{eq:b1}.
Compared with the case for 2D CMs, the critical magnetic fields are suppressed by a factor $[a/(a+l)]^2$
due to the reduction of the effective DM interaction.
The phase diagram rapidly changes at around $\tl=1$, 
where the SCoX phase and the SSA phase arise between the ferromagnetic (F) and SCyX phases
and between the SCyX and F phases, respectively [Fig.~\ref{fig:PD_CMFM}(b)].
As $\tl$ increases further, the regions of the SCyX and H phases shrink and eventually disappear for $\tl\gg 1$.
The phase boundaries among the F, SCoX, and SSA phases do not depend on $\tl$.
The dashed curve in Fig.~\ref{fig:PD_CMFM}(c) shows the F--SSA phase boundary for $\tl\gg 1$.

\section{CM/FM/CM hybrid structure}
\label{sec:CFC}
\begin{figure}[tbp]
\includegraphics[width=0.9\linewidth]{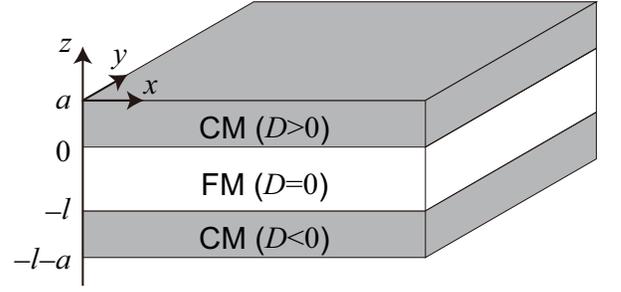}\\
\caption{Schematic of the CM/FM/CM hybrid structure,
where CM and FM denote chiral magnet and ferromagnet, respectively.
The signs of the coupling constant $D$ of the Dzyaloshinskii-Moriya interaction are opposite for two CMs.
 [See Eq.~\eqref{eq:Hamiltonian_CFC}].
}
\label{fig:CFC_sys}
\end{figure}

Next, we put another CM on the other side of the FM as shown in Fig.~\ref{fig:CFC_sys}.
We consider the case when the signs of the DM interaction in two CMs are opposite.
The total energy for this hybrid structure is given by
\begin{align}
 E=& \int_{-(a+l)}^{a} dz \iint dx dy \left[\frac{J}{2}(\nabla \nn)^2 +B(1- n_z)\right] \non\\
& - D\int_0^a  dz \nn\cdot (\nabla\times \nn)
 + D\int_{-(a+l)}^{-l} dz \nn\cdot (\nabla\times \nn).
\label{eq:Hamiltonian_CFC}
\end{align}

\begin{figure*}[tbp]
\includegraphics[width=0.99\linewidth]{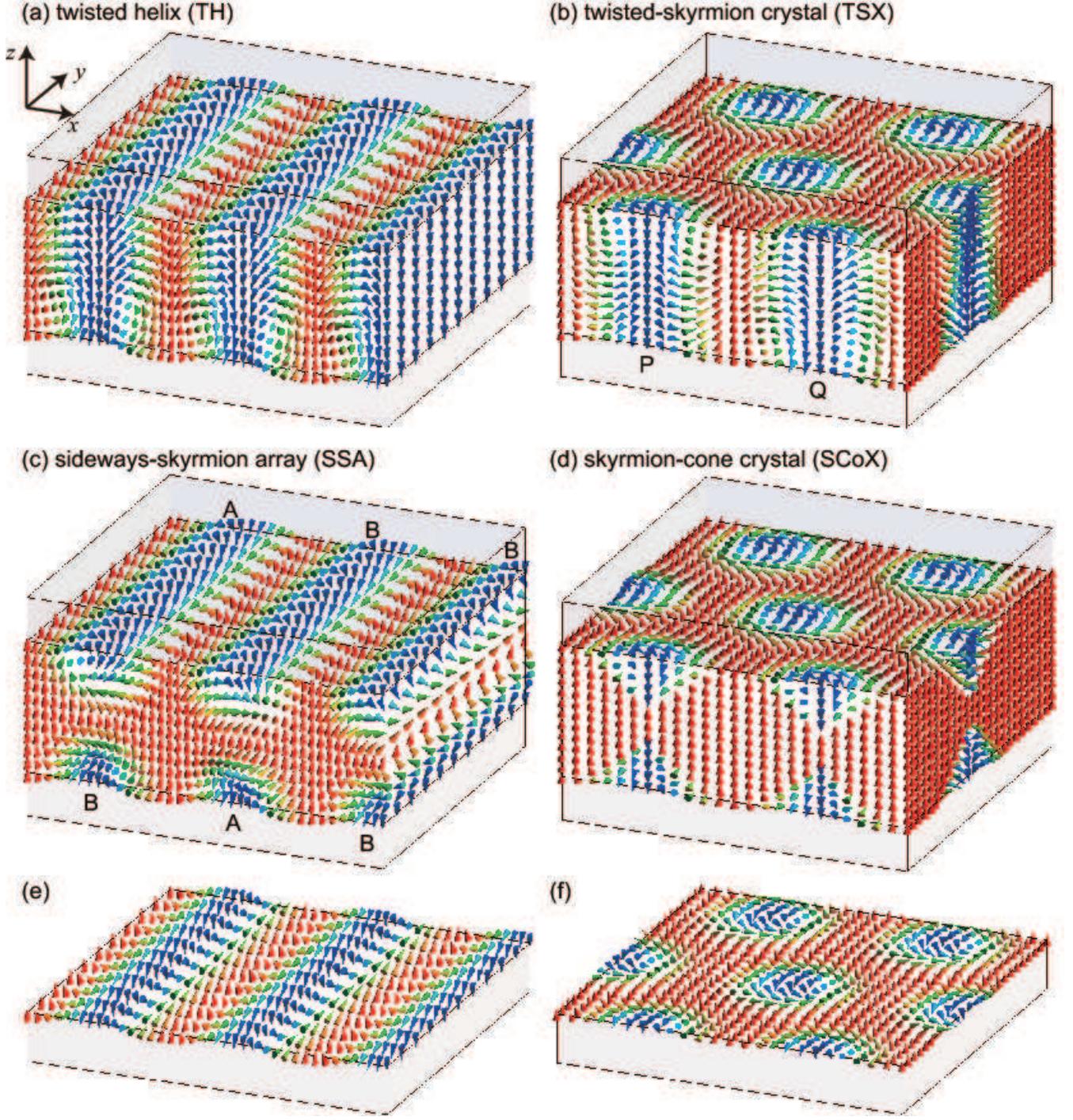}\\
\caption{(Color online) Possible three-dimensional magnetization configurations in a CM/FM/CM hybrid structure,
where CM and FM denote chiral magnet and ferromagnet, respectively.
In (a)--(d), the magnetization vectors in the FM are shown. The magnetization in the CMs, which are shown with gray cuboids, are assumed to be uniform along the $z$ direction and form spin helices [(a) and (c)] or skyrmion crystals [(b) and (d)].
The helicities of the magnetization configurations in the top and bottom CMs are opposite.
Panel (e) [(f)] shows the magnetization configuration in the bottom CM for (a) and (c) [(b) and (d)].
In (a), the helical structure is twisted by $-\pi$ about the $z$ axis as one goes from $z=0$ to $-l$ and forms a twisted helix.
Twisting by $\pi$ about the $z$ axis is also possible.
In (b), the skyrmionic structures are twisted by $\pi$ (P) or $-\pi$ (Q) about the $z$ axis as one goes from $z=0$ to $-l$. The direction of the twist is spontaneously chosen for each skyrmion.
(c) and (d) are the similar structures as Figs.~\ref{fig:CF_n} (c) and (d), respectively, where the non-uniform structures (sideways skyrmion and skyrmion cone) come in the FM from the both interfaces.
In (c), the sideways skyrmions with opposite skyrmion charges (i.e., opposite whirling patterns in the $xz$ plane) are degenerate and randomly chosen for row by row.
For the case of (c), the rows indicated by A (B) have the skyrmion charge $1/2$ ($-1/2$).
}\label{fig:CFC_n}
\end{figure*}

\begin{figure*}[tbp]
\includegraphics[width=0.99\linewidth]{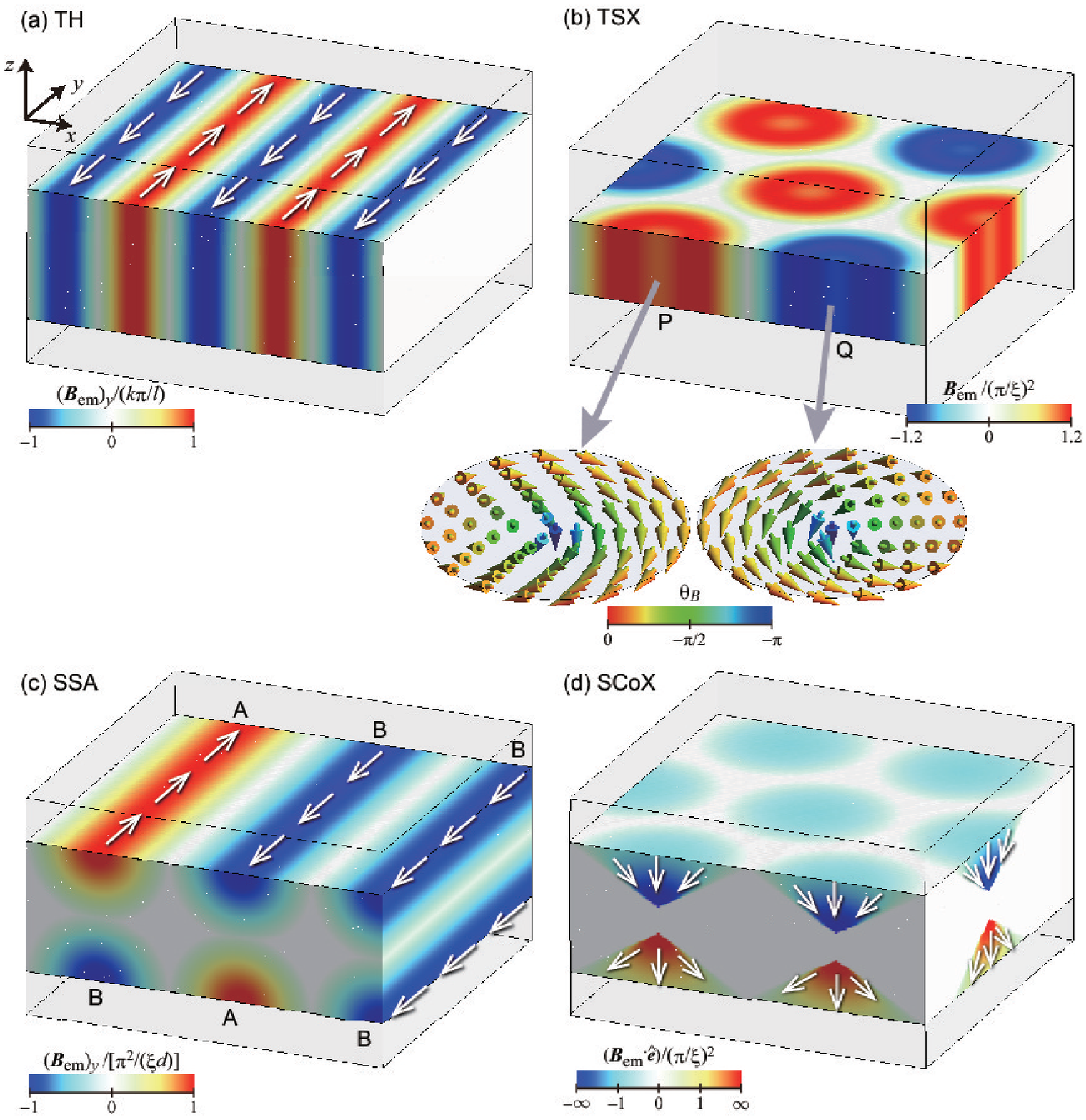}\\
\caption{(Color online) Emergent magnetic field $\bm B_{\rm em}$ for the magnetization configurations shown in Figs.~\ref{fig:CFC_n}(a)--\ref{fig:CFC_n}(d),
where TH, TSX, SSA, SCoX stand for twisted helix, twisted-skyrmion crystal, sideways-skyrmkion array, and skyrmion-cone crystal, respectively.
$\bm B_{\rm em}$ for (a), (b), (c), and (d) are given by Eqs.~\eqref{eq:Bem_TH}, \eqref{eq:Bem_TSX}, \eqref{eq:Bem_SSA}, and \eqref{eq:Bem_SCoX}, respectively.
In each panel, the color coding for the emergent magnetic field is scaled by its maximum, 
where $k$, $\xi$, and $d$ are the wave number of the helix, the skyrmion radius (or the half of the helical pitch), and the penetration depth of the non-uniform structure, respectively.
The direction of the emergent magnetic field is schematically shown with white arrows in (a), (c), and (d),
whereas that for (b) is depicted in the insets.
(a) In the TH, because of the twisting along the $z$ direction, the staggered magnetic field arises in the $y$ direction.
(b) In the TSX, the $z$ component of $\bm B_{\rm em}$ is the same as that for the SCyX shown in Fig.~\ref{fig:CF_B}(b),
but the $x$ and $y$ components, which are absent for the SCyX, arise due to the twisting.
The red (blue) color means that the whirling of the $x$ and $y$ components is clockwise (anti-clockwise) as depicted in the left (right) inset, whereas the saturation of the color shows the amplitude $|{\bm B}_{\rm em}|$.
The insets show the configuration of the emergent magnetic field in the $xy$ plane, where the colors on the arrows indicate $\theta_B\equiv \arccos[(\bm B_{\rm em})_z/|\bm B_{\rm em}|]$.
Shown are the configuration for $\tl=1$.
The skyrmions indicated by P and Q correspond to those in Fig.~\ref{fig:CFC_n}(b).
(c) The same as Fig.~\ref{fig:CF_B}(c). The rows indicated by A and B correspond to those in Fig.~\ref{fig:CFC_n}(c).
(d) The same as Fig.~\ref{fig:CF_B}(d). The red (blue) color means that the emergent magnetic field is pointing from (to) the monopole.
}\label{fig:CFC_B}
\end{figure*}

The possible magnetic structures and the resulting emergent magnetic field 
are summarized in Figs.~\ref{fig:CFC_n} and ~\ref{fig:CFC_B}, respectively.
Since the sings of the DM interactions are opposite in the two CMs, 
the helical and skyrmionic structures appearing on the top and bottom CMs have opposite helicities,
which are continuously transformed with each other by rotating the transverse magnetization by $\pm \pi$ about the $z$ axis.
Hence, a TH [Fig.~\ref{fig:CFC_n}(a)] and a TSX [Fig.~\ref{fig:CFC_n}(b)] are the possible candidates for the configuration in FM with small $l$,
which we discuss in Secs.~\ref{sec:CM/FM/CM_helix_twist} and \ref{sec:CM/FM/CM_SkX_twist}, respectively.
When $l$ becomes larger, as in the cases of the CM/FM heterostructure,
the sideways half-cylinder skyrmions and the skyrmion cones come in the FM from the CM/FM interfaces
as shown in Figs.~\ref{fig:CFC_n}(c) and \ref{fig:CFC_n}(d), respectively,
which we discuss in Sec.~\ref{sec:CM/FM/CM_SSA_SCoX}.
The phase diagram is discussed in Sec.~\ref{sec:CM/FM/CM_PD}.

\subsection{Twisted Helix}
\label{sec:CM/FM/CM_helix_twist}
We consider the following magnetic configuration:
\begin{align}
& \nn(x,y,z)  \non\\
&=\left\{
\begin{array}{ll}
\begin{pmatrix}
 0 \\ -\sin kx \\ -\cos kx
\end{pmatrix}
& (0\le z\le a),\\[7mm]
\begin{pmatrix}
 \pm\sin kx\sin(\pi z/l) \\ -\sin kx\cos(\pi z/l) \\ -\cos kx 
\end{pmatrix}
& (-l \le z\le 0),\\[7mm]
\begin{pmatrix} 
0 \\ \sin kx \\ -\cos kx
\end{pmatrix}
& (-a-l\le z\le-l).
\end{array}
\right.
\label{eq:n_TH}
\end{align}
In this magnetic profile, as we go along the $z$ direction from $z=0$ to $z=-l$, the helical texture is rotated by $\pm\pi$ about the $z$ axis
so as to continuously connect the spin helices with wave vector $k\hat{x}$ ($0\le z\le a$) and $-k\hat{x}$ ($-a-l\le z\le -l$).
The total energy for the TH is obtained by substituting Eq.~\eqref{eq:n_TH} in Eq.~\eqref{eq:Hamiltonian_CFC} as
\begin{align}
 \tE_{\rm TH}(\tk) 
&= 2\tE_{\rm hel2D}(\tk) + \pi\tl\left[\frac{\tk^2}{\AJ}+\frac{1}{2\AJ}\left(\frac{\pi}{\tl}\right)^2+\frac{b}{\AB}\right]\non\\
&=\pi(2\ta+\tl)\left[\frac{\tk^2}{\AJ} - \frac{2\ta}{2\ta+\tl}\frac{2\tk}{\AD} + \frac{b}{\AB}\right]
 + \frac{\pi^3}{2\AJ \tl}.
\label{eq:E_TH_k}
\end{align}
Comparing this with Eq.~\eqref{eq:tE_hel2D_tk}, one can immediately see that the effective DM interaction relative to the ferromagnetic and Zeeman interactions is decreased by a factor $2\ta/(2\ta+\tl)$,
and the optimized wave number and energy are respectively given by
\begin{align}
 \tk_{\rm TH} &= \frac{2\ta}{2\ta+\tl} \tk_{\rm hel2D},\\
 \tE_{\rm TH}^0
&=\pi(2\ta+\tl)\left[\frac{b}{\AB}-\left(\frac{2\ta}{2\ta+\tl}\right)^2\frac{\AJ}{\AD^2}\right]
 + \frac{\pi^3}{2\AJ \tl}.
\label{eq:E_TH}
\end{align}
The last term in the right-hand side of Eq.~\eqref{eq:E_TH} represents
the additional ferromagnetic interaction energy associated with the $z$ dependence of the magnetization profile.

The emergent magnetic field for the TH is calculated from Eqs.~\eqref{eq:n_TH} and \eqref{eq:Bem_def} as
\begin{align}
 \bm B_{\rm em} = \pm \frac{k\pi}{l}\sin kx \hat{y},
\label{eq:Bem_TH}
\end{align}
where the double sign corresponds to that in Eq.~\eqref{eq:n_TH}.
In contrast to a simple helix, which has no emergent magnetic field, the staggered magnetic field arises for the TH as shown in Fig.~\ref{fig:CFC_B}(a) due to the $z$ dependence of the magnetization configuration.

\subsection{Twisted-Skyrmion Crystal}
\label{sec:CM/FM/CM_SkX_twist}
\begin{figure}[tbp]
\includegraphics[width=0.9\linewidth]{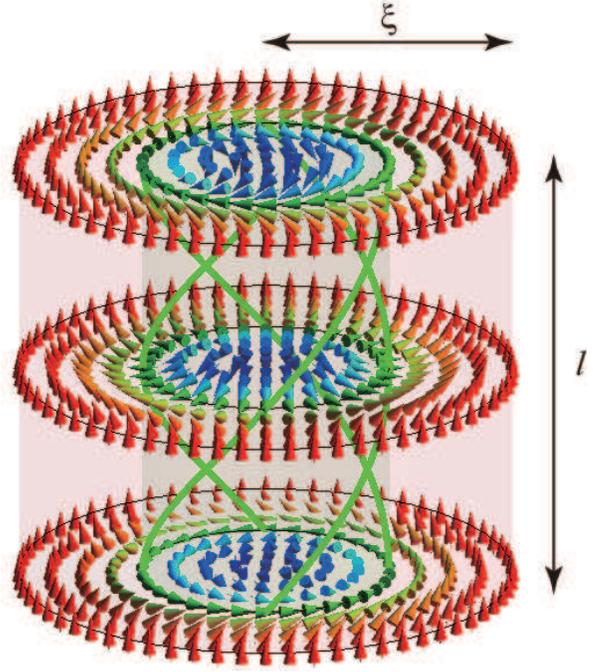}
\caption{(Color online) 
Magnetization profile in a twisted skyrmion. 
The magnetization vectors rotate by $-\pi$ about the $z$ axis as $z$ changes from $0$ to $-l$.
The green thick curves trace the positions of the same magnetization directions.
}
\label{fig:CFC_TSX2}
\end{figure}

In the case when SkXs appear in the CMs, the skyrmions in the top and bottom CMs have opposite helicities,
i.e., $\phi$ in Eq.~\eqref{eq:n_sk} is $\phi=-\pi/2 \ (\phi=\phi/2)$ for $0<z<a \ (-a-l < z < -l)$.
These structures are topologically equivalent and can be transformed to each other by a continuous transformation.
A possible structure is given by Eq.~\eqref{eq:n_sk} with taking into account the $z$ dependence of $\phi$ as shown in Fig.~\ref{fig:CFC_TSX2}.

Rewriting $\phi(z)=g(|z|/l)$ where $g(\zeta)$ is a monotonically increasing or decreasing function satisfying $g(0)=-\pi/2$ or $3\pi/2$ and $g(1)=\pi/2$,
the total energy for the crystalline structure of twisted skyrmions is given by
\begin{align}
 \tE_{\rm TSX}(\txi) =& 2\tE_{\rm SkX}(\xi)+\tl\left(\frac{1}{\txi^2}+\frac{\gamma_J}{\tl^2}+b\right)\non\\
=&(2\ta+\l)\left[\frac{1}{\txi^2}-\frac{2\ta}{2\ta+\tl}\frac{2}{\txi}+b\right] + \frac{\gamma_J}{\tl}
\label{eq:E_TSX_xi}
\end{align}
where 
\begin{align}
 \gamma_J \equiv \frac{2\pi}{\AJ }\int_0^1 d\zeta \left(\frac{dg}{d\zeta}\right)^2 \int_0^1 \rho d\rho \sin\theta^2_0(\rho),
\end{align}
and the last term of the most right-hand side of Eq.~\eqref{eq:E_TSX_xi} comes from the $z$ derivative of the magnetization.
We choose $g(\zeta)=g_+(\zeta)\equiv\pi(\zeta -1/2)$ or $g_-(\zeta)\equiv\pi(2/3-\zeta)$ and $\theta_0(\rho)=\pi(1-\rho)$, obtaining $\gamma_J=\pi^3/(2\AJ)\sim 0.40$.
Minimizing Eq.~\eqref{eq:E_TSX_xi} with respect to $\txi$, the optimized $\txi$ and the minimum energy are obtained as
\begin{align}
\txi 
&= \frac{2\ta+\tl}{2\ta},\\
\tE_{\rm TSX}^0
&= (2\ta+\tl)\left[ b - \left(\frac{2\ta}{2\ta+\tl}\right)^2\right]+\frac{\gamma_J}{\tl}.
\label{eq:E_TSX}
\end{align}
As in the case of the previous sections, the DM interaction energy relative to the other interaction energies are reduced by a factor $2\ta/(2\ta+\tl)$,
and hence, the skyrmion radius becomes larger as $\tl$ increases.
For our choice of $g(\zeta)$ and $\theta_0(\rho)$, the last term in Eq.~\eqref{eq:E_TSX}
coincides with that in Eq.~\eqref{eq:E_TH}.

The emergent magnetic field for the TSX is given by
\begin{align}
 \bm B_{\rm em} = -\left(\frac{\pi}{\xi}\right)^2\frac{\sin(\pi r/\xi)}{\pi r/\xi} \left(\mp \frac{\pi y}{l} \hat{x} \pm \frac{\pi x}{l} \hat{y}+\hat{z}\right),
\label{eq:Bem_TSX}
\end{align}
where the double sign corresponds to that of $g_{\pm}(z)$.
Here, the longitudinal component is the same as that of the SCyX shown in Fig.~\ref{fig:CF_B}(b).
However, the emergent magnetic field of the TSX also has the $x$ and $y$ components,
which are whirling in the clockwise or anti-clockwise direction depending on the direction of the twisting in the FM [see the instes of Fig.~\ref{fig:CFC_B}(b)].
Since the energies for the configurations with opposite twisting are degenerate,
the direction of the twisting is randomly chosen in each skyrmion within the framework of the variational method, as in the case of the SSA.

\subsection{Sideways-Skyrmion Array and Skyrmion-Cone Crystal}
\label{sec:CM/FM/CM_SSA_SCoX}

As in the case of the CM/FM heterostructure, the sideways skyrmions and the skyrmion cones 
may appear at the CM/FM interfaces.
The resulting structures are shown in Figs.~\ref{fig:CFC_n}(c) and \ref{fig:CFC_n}(d).
The energies for these structures are twice of those obtained in Secs.~\ref{sec:CM/FM_helix_z} and \ref{sec:CM/FM_SkX_z}.

Since the $\bm B_{\rm em}$ for a single sideways skyrmion is given by Eq.~\eqref{eq:Bem_SSA},
the emergent magnetic field for the SSA [Fig.~\ref{fig:CFC_n}(c)] is as shown in Fig.~\ref{fig:CFC_B}(c).
For the case of the CM/FM/CM hybrid system, the sideways skyrmions also appear from the bottom of the FM.
The emergent magnetic field arises in the $+\hat{y}$ or $-\hat{y}$ direction depending on the whirling direction of the magnetization vector on the $xz$ plane,
which is randomly chosen for row by row.

The emergent magnetic field for the SCoX [Fig.~\ref{fig:CFC_n}(d)] is shown in Fig.~\ref{fig:CFC_B}(d).
As in the case of Fig.~\ref{fig:CF_B}(d), the emergent magnetic field diverges as one approaches the monopole.
The crucial difference from Fig.~\ref{fig:CF_B}(d) is, however, that the direction of the emergent magnetic field is dependent on which interface the skyrmion cone comes out from:
For the skyrmion cone coming out from the top (bottom) CM/FM interface, 
the  emergent magnetic field points to (from) the monopole on the top of the cone.

\subsection{Phase Diagram}
\label{sec:CM/FM/CM_PD}
\begin{figure}[tbp]
\includegraphics[width=0.9\linewidth]{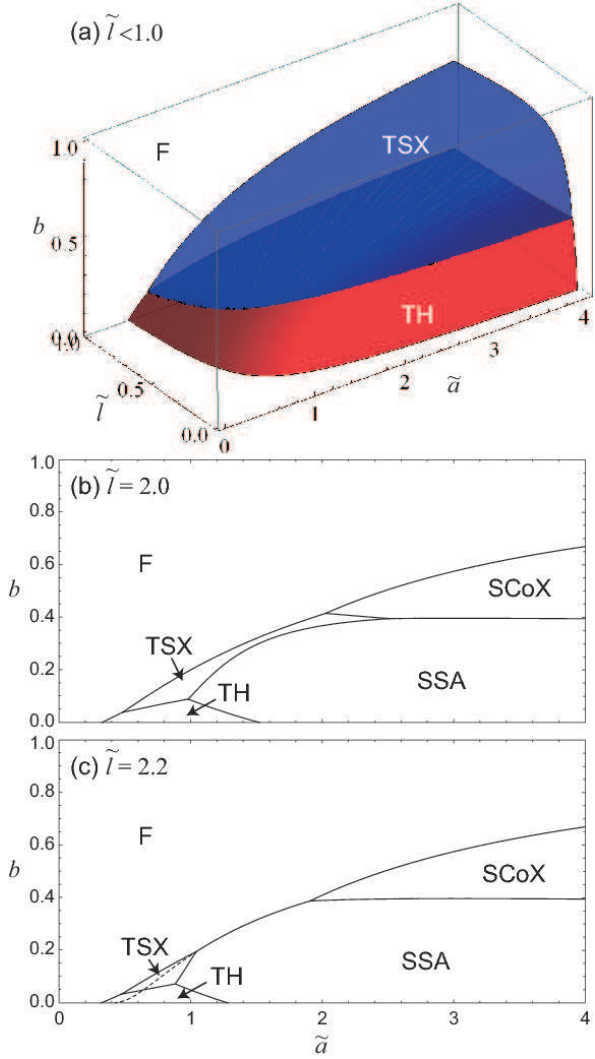}
\caption{(Color online) Phase diagram of the magnetization structures shown in Fig.~\ref{fig:CFC_n}
in a CM/FM/CM hybrid structure with (a) $\tl=1.0$, (b) $\tl=2.0$, and (c) $\tl=2.2$,
where F, TSX, TH, SCoX, SSA stand for ferromagnetic, twisted-skyrmion crystal, twisted helix, skyrmion-cone crystal, and sideways-skyrmion array phases, respectively.
The F--TH, F--TSX and TH--TSX phase boundaries are given by
Eqs.~\eqref{eq:b_F-TH}, \eqref{eq:b_F-TSX}, and \eqref{eq:b_TH-TSX}, respectively,
whereas the other phase boundaries are numerically calculated.
The F--SCoX, F--SSA, and SCoX--SSA phase boundaries are the same as those for the CM/FM system and independent of $\tl$.
The dashed curve in (c) indicates the F--SSA phase boundary for $\tl\gg1$.
}
\label{fig:PD_CMFMCM}
\end{figure}

By compareing the energy for each configuration, 
we obtain the phase diagram of the CM/FM/CM hybrid system as shown in Fig.~\ref{fig:PD_CMFMCM},
where we calculate for (a) $\tl< 1.0$, (b) $\tl=2.0$, and (c) $\tl=2.2$.
When $\tl<1.0$, only the TSX, TH, and F phases appear.
The F--TH, F--TSX and TH--TSX phase boundaries are given by
\begin{align}
b_\textrm{F-TH}&=\frac{\AB \AJ}{\AD^2}\left(\frac{2\ta}{2\ta+\tl}\right)^2 -\frac{\gamma_J\AB}{\pi\tl(2\ta+\tl)},
\label{eq:b_F-TH}\\
b_\textrm{F-TSX}&=\left(\frac{2\ta}{2\ta+\tl}\right)^2-\frac{\gamma_J}{\tl(2\ta+\tl)},
\label{eq:b_F-TSX}\\
b_\textrm{TH-TSX}&=\left(\frac{2\ta}{2\ta+\tl}\right)^2 b_1,
\label{eq:b_TH-TSX}
\end{align}
respectively, where we have used $\gamma_J=\pi^3/(2\AJ)$, and $b_1$ is defined in Eq.~\eqref{eq:b1}.
In contrast to the case of Fig.~\ref{fig:PD_CMFM}(a), where the H and SCyX phases start from $\ta=0$, there is a lower bound of $\ta$ for the appearance of the TH and TSX phases.
For example, From Eq.~\eqref{eq:b_F-TH}, the F--TH phase boundary at $b=0$ is given by
\begin{align}
 \ta = \frac{\pi\AD^2}{4\AJ\tl}\left(\frac{\pi}{2\AJ}+\sqrt{\frac{\pi^2}{4\AJ^2}+\frac{2\tl^2}{\AD^2}}\right),
\end{align}
which behaves as 
\begin{align}
 \ta &\sim \left(\frac{\pi\AD}{2\AJ}\right)^2\frac{1}{\tl} \ \ \ (\tl \to 0),\\
 \ta &\sim \frac{\pi\AD}{2\sqrt{2}\AJ} \sim 0.28 \ \ \ (\tl\to\infty).
\end{align}
This is because the ferromagnetic interaction energy associated with the $z$ dependence of the magnetization structure prevents the system
from creating non-uniform structure.
In order to overcome the energy cost in the FM, the thickness of the CMs, which have the negative DM interaction energy, should be large enough.

As $\tl$ increases [Fig.~\ref{fig:PD_CMFMCM}(b)],
the SCoX and SSA phases arise between the F and TSX phases and between the TH and TSX phases, respectively, 
and the regions of the TH and TSX phases rapidly shrinks [Fig.~\ref{fig:PD_CMFMCM}(c)].

\section{Discussion and conclusion}
\label{sec:discussion}

We have discussed possible magnetization configurations in ground states at CM/FM and CM/FM/CM hybrid structures.
The energy of the system is calculated by using a variational method, where
we assume a certain magnetization structures and take its length scales, i.e.,
the skyrmion radius and the penetration depth, as variational parameters.
By comparing the obtained energies, the ground-state phase diagrams of CM/FM and CM/FM/CM hybrid structures
are obtained as shown in Figs.~\ref{fig:PD_CMFM} and \ref{fig:PD_CMFMCM}, respectively,
where 3D exotic configurations, such as SSA, SCoX, TH, and TSX, appear in low magnetic fields.

In particular, the interface introduces a sort of frustration and hence can produce nontrivial magnetization textures absent in each constitute alone. For example, both helix and ferromagnet are not topological, while the SSA which emerges at the interface between these two is topological characterized by the emergent magnetic field. The phase diagrams in Figs.~\ref{fig:PD_CMFM} and \ref{fig:PD_CMFMCM} will provide a basis to design these nontrivial magnetization structures in the interface systems.

Transport properties of conduction electrons coupled to the magnetization is greatly influenced by the emergent electromagnetic field. The distribution of the emergent magnetic field $\bm{B}_{\rm em}$ shown in Figs.~\ref{fig:CF_B} and \ref{fig:CFC_B} will produce various topological Hall effects depending on the direction of $\bm{B}_{\rm em}$. Especially, the diverging $\bm{B}_{\rm em}$ at the monopole is expected to affect the electron motion strongly. (Note that the lattice constant gives the cut-off for this divergence in real systems.) Furthermore, the current-driven motion of the magnetization textures via the spin transfer torque is determined by the gyro-vector $\bm{G}$, i.e., the integral of $\bm{B}_{\rm em}$ over the space. $\bm{G}$ enters into the Thiele's equation and the finite $\bm{G}$ enhances the spin transfer torque effect~\cite{Iwasaki2013a}. Once the current-driven motion of $\bm{B}_{\rm em}$ occurs, the emergent electric field $\bm{E}_{\rm em}$ is induced, i.e., emergent electromagnetic induction. The design of these varieties of phenomena in the heterostructures will open a rich physics of magnetic textures.

\acknowledgments
This work was supported by Grants-in-Aid for Scientific
Research (No. 22740265, No. 15K17726, No. 26103006, Kiban S No. 24224009, and Research on Innovative Areas "Topological Materials Science" No.15H05853) 
from the Ministry of Education, Culture, Sports, Science and Technology (MEXT) of Japan and from Japan Society for the Promotion of Science.

%

\end{document}